\documentstyle[dcolumn,preprint,epsf,aps,floats,tighten]{revtex}

\begin{document}
\lefthyphenmin=2
\righthyphenmin=3
\title{\boldmath Small Angle Muon and Bottom Quark Production in  
$p \overline{p}$\ Collisions at $\sqrt{s} = 1.8$ TeV}
                                                                     

%
\author{                                                                      
B.~Abbott,$^{45}$                                                             
M.~Abolins,$^{42}$                                                            
V.~Abramov,$^{18}$                                                            
B.S.~Acharya,$^{11}$                                                          
I.~Adam,$^{44}$                                                               
D.L.~Adams,$^{54}$                                                            
M.~Adams,$^{28}$                                                              
S.~Ahn,$^{27}$                                                                
V.~Akimov,$^{16}$                                                             
G.A.~Alves,$^{2}$
% ...
\\                                                                            
\vskip 0.20cm                                                                 
\centerline{(D\O\ Collaboration)}                                             
\vskip 0.20cm                                                                 
}                                                                             
\address{                                                                     
\centerline{$^{1}$Universidad de Buenos Aires, Buenos Aires, Argentina}       
\centerline{$^{2}$LAFEX, Centro Brasileiro de Pesquisas F{\'\i}sicas,         
                  Rio de Janeiro, Brazil}                                     
\centerline{$^{3}$Universidade do Estado do Rio de Janeiro,                   
                  Rio de Janeiro, Brazil}                                     
\centerline{$^{4}$Institute of High Energy Physics, Beijing,                  
                  People's Republic of China}                                 
\centerline{$^{5}$Universidad de los Andes, Bogot\'{a}, Colombia}             
\centerline{$^{6}$Universidad San Francisco de Quito, Quito, Ecuador}         
\centerline{$^{7}$Institut des Sciences Nucl\'eaires, IN2P3-CNRS,             
                  Universite de Grenoble 1, Grenoble, France}                 
\centerline{$^{8}$DAPNIA/Service de Physique des Particules, CEA, Saclay,     
                  France}                                                     
\centerline{$^{9}$Panjab University, Chandigarh, India}                       
\centerline{$^{10}$Delhi University, Delhi, India}                            
\centerline{$^{11}$Tata Institute of Fundamental Research, Mumbai, India}     
\centerline{$^{12}$Kyungsung University, Pusan, Korea}                        
\centerline{$^{13}$Seoul National University, Seoul, Korea}                   
\centerline{$^{14}$CINVESTAV, Mexico City, Mexico}                            
\centerline{$^{15}$Institute of Nuclear Physics, Krak\'ow, Poland}            
\centerline{$^{16}$Institute for Theoretical and Experimental Physics,        
                   Moscow, Russia}                                            
\centerline{$^{17}$Moscow State University, Moscow, Russia}                   
\centerline{$^{18}$Institute for High Energy Physics, Protvino, Russia}       
\centerline{$^{19}$Lancaster University, Lancaster, United Kingdom}           
\centerline{$^{20}$University of Arizona, Tucson, Arizona 85721}              
\centerline{$^{21}$Lawrence Berkeley National Laboratory and University of    
                   California, Berkeley, California 94720}                    
\centerline{$^{22}$University of California, Davis, California 95616}         
\centerline{$^{23}$University of California, Irvine, California 92697}        
\centerline{$^{24}$University of California, Riverside, California 92521}     
\centerline{$^{25}$Florida State University, Tallahassee, Florida 32306}      
\centerline{$^{26}$University of Hawaii, Honolulu, Hawaii 96822}              
\centerline{$^{27}$Fermi National Accelerator Laboratory, Batavia,            
                   Illinois 60510}                                            
\centerline{$^{28}$University of Illinois at Chicago, Chicago,                
                   Illinois 60607}                                            
\centerline{$^{29}$Northern Illinois University, DeKalb, Illinois 60115}      
\centerline{$^{30}$Northwestern University, Evanston, Illinois 60208}         
\centerline{$^{31}$Indiana University, Bloomington, Indiana 47405}            
\centerline{$^{32}$University of Notre Dame, Notre Dame, Indiana 46556}       
\centerline{$^{33}$Purdue University, West Lafayette, Indiana 47907}          
\centerline{$^{34}$Iowa State University, Ames, Iowa 50011}                   
\centerline{$^{35}$University of Kansas, Lawrence, Kansas 66045}              
\centerline{$^{36}$Kansas State University, Manhattan, Kansas 66506}          
\centerline{$^{37}$Louisiana Tech University, Ruston, Louisiana 71272}        
\centerline{$^{38}$University of Maryland, College Park, Maryland 20742}      
\centerline{$^{39}$Boston University, Boston, Massachusetts 02215}            
\centerline{$^{40}$Northeastern University, Boston, Massachusetts 02115}      
\centerline{$^{41}$University of Michigan, Ann Arbor, Michigan 48109}         
\centerline{$^{42}$Michigan State University, East Lansing, Michigan 48824}   
\centerline{$^{43}$University of Nebraska, Lincoln, Nebraska 68588}           
\centerline{$^{44}$Columbia University, New York, New York 10027}             
\centerline{$^{45}$New York University, New York, New York 10003}             
\centerline{$^{46}$University of Rochester, Rochester, New York 14627}        
\centerline{$^{47}$State University of New York, Stony Brook,                 
                   New York 11794}                                            
\centerline{$^{48}$Brookhaven National Laboratory, Upton, New York 11973}     
\centerline{$^{49}$Langston University, Langston, Oklahoma 73050}             
\centerline{$^{50}$University of Oklahoma, Norman, Oklahoma 73019}            
\centerline{$^{51}$Brown University, Providence, Rhode Island 02912}          
\centerline{$^{52}$University of Texas, Arlington, Texas 76019}               
\centerline{$^{53}$Texas A\&M University, College Station, Texas 77843}       
\centerline{$^{54}$Rice University, Houston, Texas 77005}                     
}                                                                             

\maketitle
\vspace{-0.8cm}
\begin{abstract}
\vskip -1.0cm
{
This Letter describes a measurement of the muon cross section originating 
from $b$ quark decay in the forward rapidity range 
$2.4<|{\rm \sl y}^{\mu}|<3.2$ in
$p\bar{p}$ collisions at $\sqrt{s}$ = 1.8 TeV. The data used in this analysis
were collected by the D\O\ experiment at the Fermilab Tevatron. We find that
NLO QCD calculations underestimate $b$ quark production
by a factor of four in the forward rapidity region.
}
\end{abstract}



Measurements of $b$ quark production at the Tevatron have provided 
valuable information in the study of perturbative QCD. 
Cross sections measured by both the D\O\ \cite{ref:DO_bx,ref:DO_dim},  and 
CDF \cite{ref:CDF_bx} collaborations in the central rapidity region
 ($|{\rm \sl y}^{b}|<1.5$) are
systematically higher (by a factor of two to 
three) than the nominal values predicted by next-to-leading order
(NLO) QCD \cite{ref:MNR}. 
Furthermore, discrepancies between some of the predicted and measured
shapes of $b\bar{b}$ correlation distributions \cite{ref:CDF_corr} indicate
that the difference between data and QCD cannot be explained by a simple
normalization factor. 

CDF has recently measured
the $b{\bar b}$ cross section in which 
one quark is produced in the forward pseudorapidity region
($1.8 < | \eta^{b} | < 2.6$) and the
other in the central range
($| \eta^{b} | < 1.5$)\cite{ref:CDF_forw}. This measurement is
a factor of 2.4 higher than the NLO QCD prediction.

Our measurement of the forward cross section of muons originating from
$b$ quark decay extends these studies to the previously unexplored
rapidity region ($2.4 < | {\rm \sl y}^{\mu} | < 3.2$), and
provides further insights into the discrepancy between
$b$ quark production measurements and theoretical predictions.

Forward muons are measured by the D\O\ detector \cite{ref:D0_det} using the
Small Angle MUon Spectrometer (SAMUS) \cite{ref:sam_det,ref:sampsi}. 
SAMUS consists of two identical systems, each with three drift tube stations
and a 1.8~T magnetized iron toroid, on either side of the interaction region.
The momentum  resolution of this system varies from $\approx 19\%$ at 20 GeV/$c$
to $\approx 25\%$ at 100 GeV/$c$.
Muons reaching the SAMUS chambers traverse approximately 20 interaction lengths
of material, reducing the hadronic punch-through background to a
negligible level.
This region does, however, face a large combinatoric
background  due to the flux of beam
jet related particles.
There are on average  6 to 14  hits per plane in a given
bunch crossing, and the drift tubes near the beam axis have
an approximate $5\%$  occupancy. 

The data used in this analysis come from special runs taken at low instantaneous
luminosity during the 1994-95 collider run.  The integrated luminosity for these
runs is $104 \pm 6 \;{\rm nb}^{-1}$. The trigger 
required the presence of an inelastic collision near the center of the
detector and at least one track in the SAMUS detector 
with an apparent $p_T^{\mu} >3$ GeV/$c$ pointing back to
the interaction region. Muon candidates were also required to have an 
associated energy deposition in the calorimeter. 
The hit multiplicity in each layer was also required to fall below a maximum
cutoff to improve background rejection and 
lower the trigger rates to an acceptable level. 

Muons are selected offline in the rapidity range
$2.4 <|{\rm \sl y}^{\mu}|<3.2$, with $p^{\mu}<150$ GeV/$c$ 
and $p_T^{\mu} > 2$ GeV/$c$.
Single interaction events are selected by requiring only one
reconstructed vertex in an event,
leaving an effective integrated luminosity of
${\cal L} = 75\pm 7 \; {\rm nb}^{-1}$. Muon tracks are required to
have at least 15, out of an average of 18, hits.
To ensure a good momentum measurement, we require muons to traverse a 
magnetic field integral of at least 1.2~T$\cdot$m.
Muons are also required to be associated with a 
track-like object in the calorimeter with energy deposition consistent with
that of a minimum ionizing particle. With these cuts, the combinatoric background
is determined using both data and Monte Carlo (MC) to be less that 1\%.
The number of surviving muons in this sample is $N^\mu = 5106$.

The muon trigger and track reconstruction efficiencies
are obtained using data and MC single muons, with detector simulation
using {\footnotesize GEANT} \cite{ref:GEANT}, superimposed onto
real minimum bias events. The trigger efficiencies for the hit multiplicity cut
[($31 \pm 2)\%$] and the calorimeter confirmation [($95 \pm 1)\%$] 
are obtained from data, as are the offline cut efficiencies for 
energy deposition [($94 \pm 3)\%$] and number of hits on a track [($96 \pm 2)\%$].
The overall detection efficiency is $1\%$ for $p_T^{\mu}$ = 2 GeV/$c$ and
reaches a plateau of $10\%$ for $p_T^{\mu}>9$ GeV/$c$.
The MC momentum scale and resolution are shown to be correct to
within $2\%$ by comparing the peak values and widths of the reconstructed
$J/\psi$ signal from data \cite{ref:sampsi} and MC.  

\begin{figure}[t]
\vspace{-0.5cm}
\centerline{\epsfxsize=7in\epsfbox{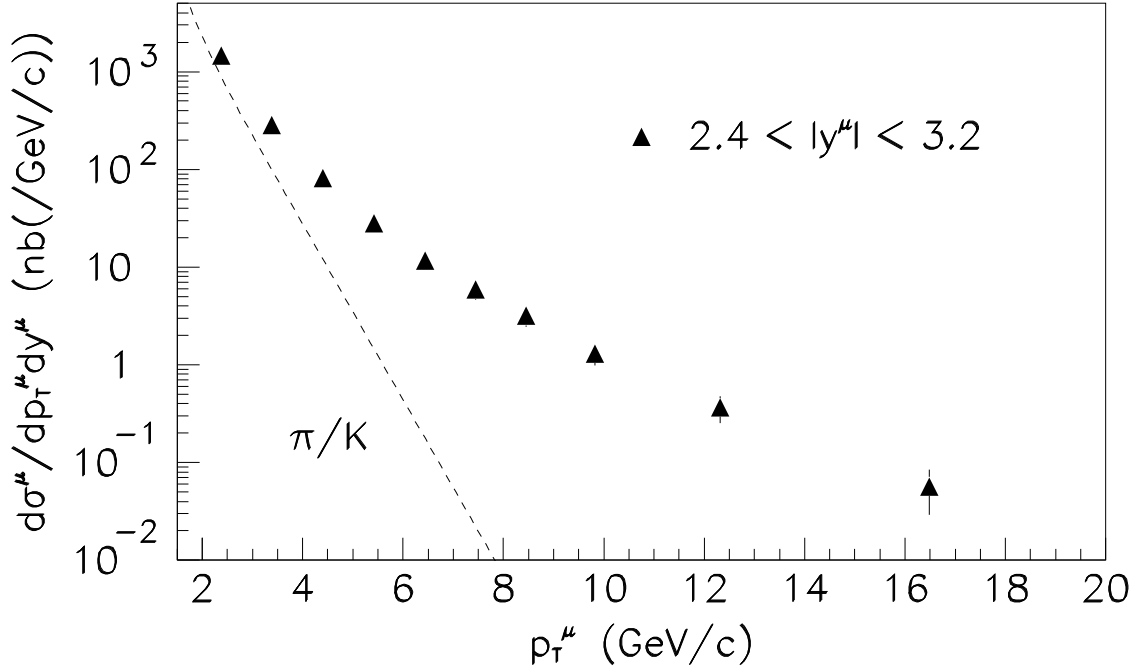}}
\caption{The inclusive muon cross section in the
forward region as a function of $p_T^{\mu}$  (per unit rapidity). The
dashed line shows the expected contributions from $\pi/K$ decays.}
\label{fig:mu_xs}
\end{figure}

The muon cross section is calculated
as follows:
\begin{equation}
\scriptsize
{\frac{d\sigma^{\mu}}{dp_{T}^{\mu}d{\rm \sl y}^{\mu}}} = {\frac{1}{{\cal L} 
\Delta {\rm \sl y}^{\mu} \Delta p_T^{\mu}}}
{\frac{N^{\mu} f_{\rm smr}}{\epsilon}},
\large
\end{equation}
where $f_{\rm smr}$ is a correction factor that accounts for momentum smearing,
and $\epsilon$ is the detection efficiency.
As there are high correlations between kinematic variables and cuts, 
$f_{\rm smr}$ and $\epsilon$ are determined by
\begin{equation}
\scriptsize
{\frac{N^{\mu} f_{\rm smr}}{\epsilon}}={\frac{1}{\epsilon_{\rm data}}}{\frac{H({\rm data})
H({\rm MCgen})} {H({\rm MCreco})}},
\large
\end{equation}
where $\epsilon_{\rm data}$ is the combined data-based efficiency of the
previously described cuts not simulated in the MC,  and the $H$'s are 
matrices with elements corresponding to two-dimensional
histograms in the $(p_T^{\mu}, {\rm \sl y}^{\mu})$ plane. $H({\rm data})$ is the data
distribution after all offline cuts; $H({\rm MCgen})$ is the generated Monte 
Carlo distribution, and $H({\rm MCreco})$ is the reconstructed MC
distribution with full detector simulation and the same cuts as the data.
The histograms are segmented with 25
bins in $p_T^{\mu}$ from 0 to 25 GeV/$c$, and 7 bins in rapidity from 2.0 to 3.4.
The MC events are weighted in an iterative procedure to match the
corrected $p_T^{\mu}$ and rapidity distributions of the data. 
This method is found to give consistent results (within $3\%$) regardless
of the shape of the initial distribution.
The resulting reconstructed MC distributions also agree quite well with
those of the data for all kinematic variables of interest after the weighting
procedure.

The inclusive muon cross section in the forward rapidity region (which includes
both muon charges)
is shown in Fig.~\ref{fig:mu_xs} and Table~\ref{tab:bmu_xs}. The systematic
errors in this measurement vary as a function of $p_T^{\mu}$ from 
15 to $45\%$. They are dominated by uncertainties associated with the
momentum smearing correction [$(6-41)\%$],
the single interaction luminosity $(10\%)$, and the trigger efficiency
$(8\%)$.

\newcolumntype{3}{D{.}{.}{4.3}}
\newcolumntype{2}{D{.}{.}{2.3}}
\newcolumntype{1}{D{.}{.}{1.3}}
\newcolumntype{4}{D{.}{.}{1.4}}
\newcolumntype{-}{D{-}{-}{3.4}}
\newcolumntype{5}{D{.}{.}{4.2}}
\newcolumntype{6}{D{.}{.}{2.3}}
\newcolumntype{7}{D{.}{.}{3.2}}
\newcolumntype{8}{D{.}{.}{3.3}}
\newcolumntype{9}{D{.}{.}{2.1}}
\footnotesize
\begin{table*}[t] 
\centering
\caption{Forward muon cross sections (per unit rapidity).}
\vskip 0.3cm
\label{tab:bmu_xs}
\begin{tabular}{c|9|3c6c8|5c7|6c4|2c1c2}
\hline
\hline
\multicolumn{1}{c|}{$p_T^{\mu}$} & \multicolumn{1}{c|}{$ \langle p_T^{\mu} \rangle$} & \multicolumn{5}{c|}{$\sigma^{\mu}$}
& \multicolumn{3}{c|}{$\sigma^{\mu} (\pi/K)$} & \multicolumn{3}{c|}{$f_b$} & \multicolumn{5}{c}{$\sigma^{\mu}_{b}$} \\
 
\multicolumn{1}{c|}{(GeV/$c$)} & \multicolumn{1}{c|}{(GeV/$c$)} & \multicolumn{5}{c|}{(nb/(GeV/$c$))}
&\multicolumn{3}{c|}{(nb/(GeV/$c$))} & & & & \multicolumn{5}{c}{(nb/(GeV/$c$))} \\
\hline
2 -- 3   & 2.4  & 1474  & $\pm$ & 33    & $\pm$ & 265   & 1091 & $\pm$ & 383  &       &       &       &       &       &       &       &    \\
3 -- 4   & 3.4  & 282.5 & $\pm$ & 7.5   & $\pm$ & 45    & 92.2 & $\pm$ & 33.1 & 0.513 & $\pm$ & 0.087 & 97.6  & $\pm$ & 3.8   & $\pm$ & 25 \\
4 -- 5   & 4.4  & 81.4  & $\pm$ & 3.1   & $\pm$ & 12    & 10.4 & $\pm$ & 3.7  & 0.619 & $\pm$ & 0.086 & 43.9  & $\pm$ & 1.9   & $\pm$ & 9.2 \\
5 -- 6   & 5.4  & 28.2  & $\pm$ & 1.5   & $\pm$ & 4.2   & 1.3  & $\pm$ & 0.5  & 0.656 & $\pm$ & 0.078 & 17.6  & $\pm$ & 1.0   & $\pm$ & 3.4 \\
6 -- 7   & 6.4  & 11.72 & $\pm$ & 0.80  & $\pm$ & 1.9   & 0.17 & $\pm$ & 0.06 & 0.671 & $\pm$ & 0.080 & 7.75  & $\pm$ & 0.54  & $\pm$ & 1.6 \\
7 -- 8   & 7.4  & 5.86  & $\pm$ & 0.53  & $\pm$ & 1.1   & 0.02 & $\pm$ & 0.01 & 0.675 & $\pm$ & 0.081 & 3.94  & $\pm$ & 0.36  & $\pm$ & 0.83 \\
8 -- 9   & 8.4  & 3.17  & $\pm$ & 0.34  & $\pm$ & 0.63  &      &       &      & 0.685 & $\pm$ & 0.075 & 2.17  & $\pm$ & 0.23  & $\pm$ & 0.50 \\
9 -- 11  & 9.8  & 1.30  & $\pm$ & 0.13  & $\pm$ & 0.29  &      &       &      & 0.697 & $\pm$ & 0.070 & 0.906 & $\pm$ & 0.091 & $\pm$ & 0.22 \\
11 -- 15 & 12.4  & 0.367 & $\pm$ & 0.039 & $\pm$ & 0.11  &      &       &      & 0.718 & $\pm$ & 0.067 & 0.264 & $\pm$ & 0.028 & $\pm$ & 0.080 \\
15 -- 20 & 16.7  & 0.057 & $\pm$ & 0.011 & $\pm$ & 0.026 &      &       &      & 0.749 & $\pm$ & 0.062 & 0.043 & $\pm$ & 0.008 & $\pm$ & 0.020\\
\hline
\hline
\end{tabular}
\end{table*}
\normalsize

The contributions to this cross section
from cosmic rays, hadronic punch-through, and $W$/$Z$ decay are
negligible (determined using both data and MC). 
The pion and kaon decay contribution is obtained using
{\footnotesize ISAJET} \cite{ref:ISAJET}, which we find to
be in agreement with the charged particle cross section 
measured in the central region \cite{ref:CDF_pik}.
The excess above the $\pi/K$ contribution is attributed 
to $b$ and $c$ quark decay. The fraction 
of this excess due to $b$ quark decay ($f_b$) can be obtained
using the transverse momentum spectrum of the muons relative to that
of an associated jet ($p_T^{\rm rel}$), 
but, because of our jet 
reconstruction threshold of $E_T > 10$ GeV, only $(7.9 \pm 0.8 )\%$ of the
events in the forward region have a reconstructed associated jet.
We must, therefore, rely on a NLO QCD MC to determine $f_b$.


In this Monte Carlo, $b$ and $c$ quarks are
generated according to the $p_T$ and rapidity distributions of NLO QCD
calculations \cite{ref:MNR} using MRSR2 parton distribution functions
\cite{ref:MRS}, quark masses $m_{b}$ = 4.75 GeV/$c^2$ 
and $m_{c}$ = 1.6 GeV/$c^2$, with renormalization and factorization scales
$\mu=\mu_{0}=\sqrt{m_{q}^{2}+p_{T}^{2}}$.
The four-momenta of the quarks are input to an {\footnotesize ISAJET}
MC which simulates initial and final state radiation, as well as quark
fragmentation and decay. The theoretical uncertainty is determined by varying
the parameters $m_b$ from 4.5 to 5.0 GeV/$c^2$, 
$m_c$ from 1.3 to 1.9 GeV/$c^2$, and $\mu$ from $\mu_{0}/2$ to $2\mu_{0}$.
The Peterson fragmentation
parameters \cite {ref:Peterson}($\epsilon_b = 0.006$, $\epsilon_c = 0.06$)
are also varied by $50\%$, as are the branching ratios within their
errors \cite{ref:PDG}. This simulation
predicts that $8.5 \%$ of the muons should have a reconstructed associated
jet, which is consistent within errors with what is found in the data. 

\begin{figure}[t]
\epsfxsize=7.0in\epsfbox{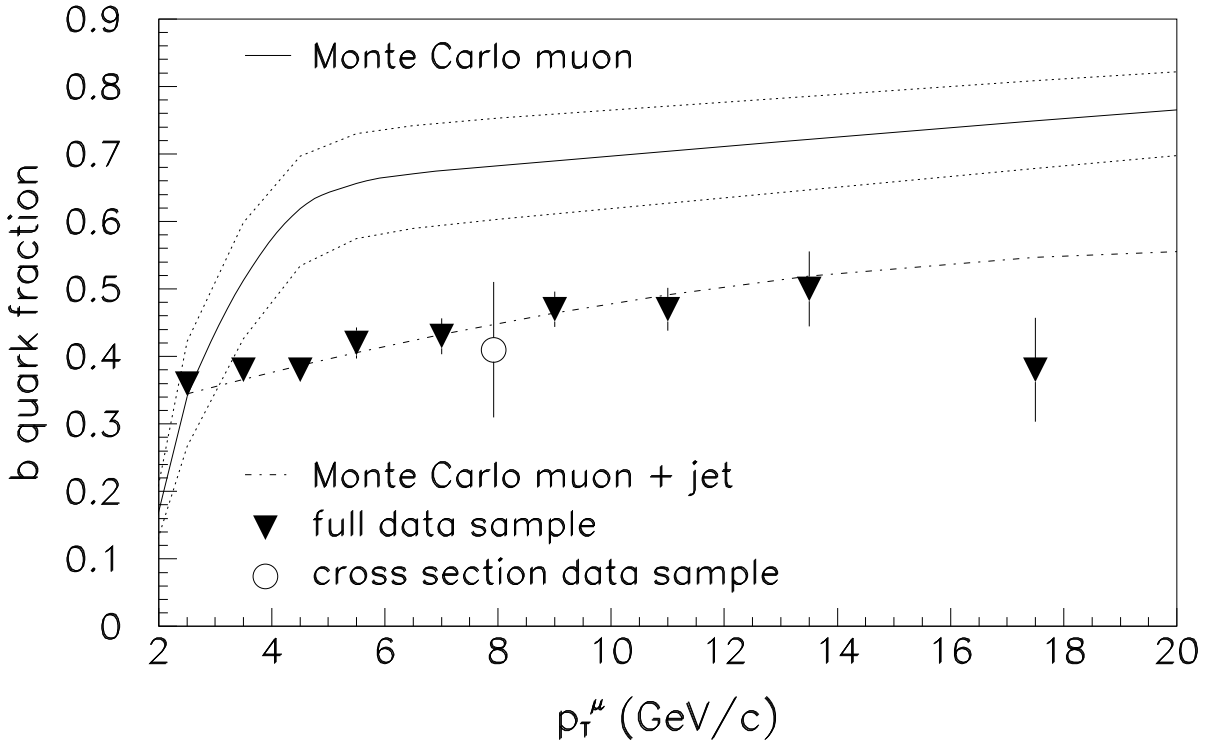}
\caption {$f_b$ for muons with 
an associated jet as measured from data $p_T^{\rm rel}$ fits
(triangles and circle) and as predicted by the NLO QCD MC (dot-dashed curve)
The prediction of $f_b$ for muons without the jet requirement is shown by
the solid curve with uncertainties indicated by dotted curves.}
\label{fig:b_fracts}
\end{figure}

We check the validity of this MC by comparing its prediction
for $f_b$ to that determined from our entire 1994-95 data set.
Thirty-one thousand forward muons with an associated jet are selected
from low $p_T$ single muon and muon+jet triggers. The trigger requirements
keep the physics content of this sample the same as that of the cross section
sample. The full sample is unsuitable for a cross section determination, however,
as there is a large uncertainty in its normalization due to the various trigger
thresholds and pre-scales, and luminosities that the data was taken with.
\begin{figure}[t]
\epsfxsize=7in\epsfbox{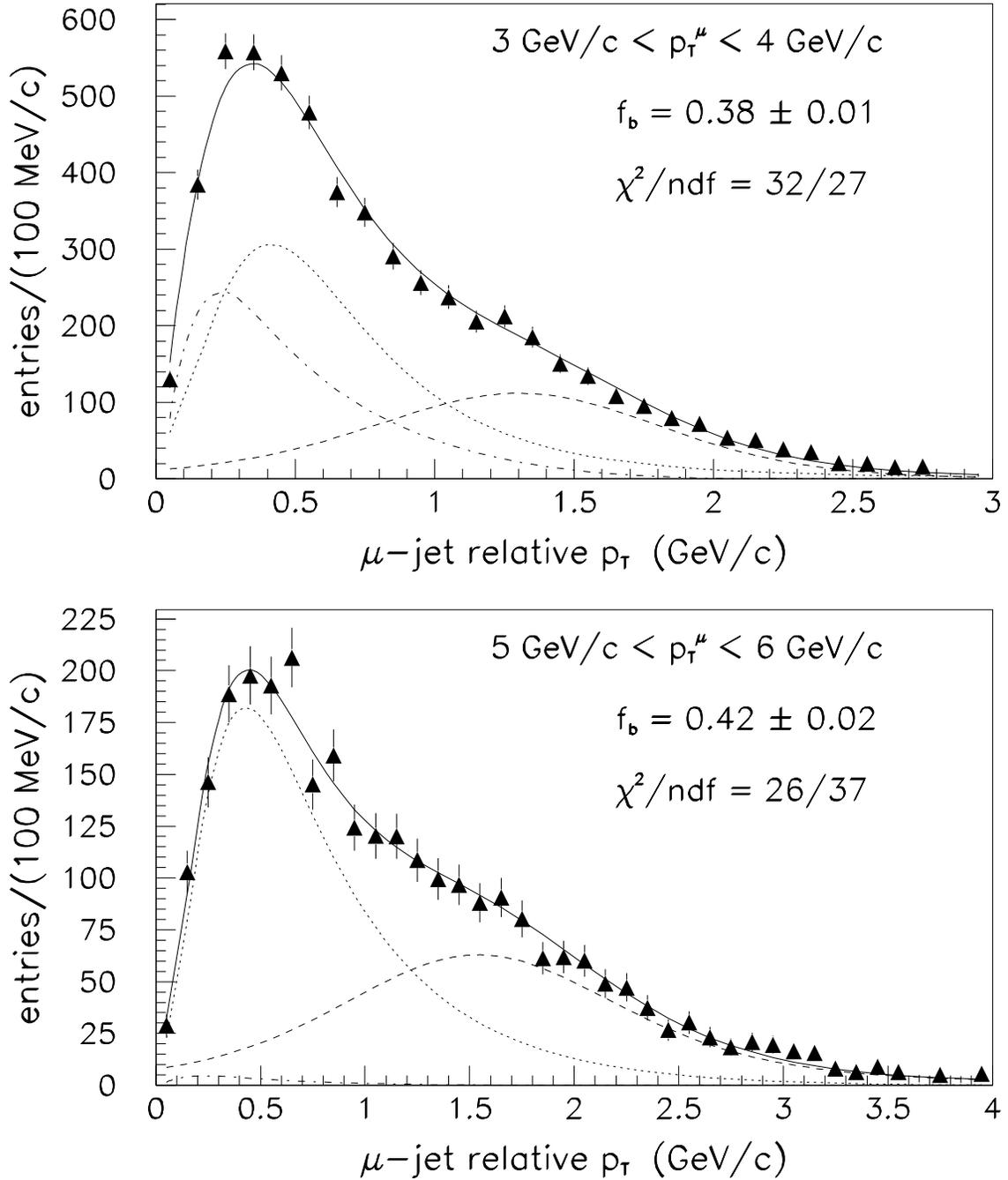}
\caption {Data $p_T^{\rm rel}$ distributions for two selected $p_T^{\mu}$ ranges. 
The solid line shows the fit to the data, with broken lines showing
contributions from $b$ quark (dashed), $c$ quark (dotted),
and $\pi /K$ (dot-dashed) decay. $f_b$ is the $b$ quark fraction
after $\pi /K$ subtraction (errors are statistical only).}  
\label{fig:ptrel}
\end{figure}
The $b$ quark fraction is determined by fitting
the $p_T^{\rm rel}$ distributions (in various ranges of $p_T^{\mu}$) to the
expected shapes from $b$ quark, $c$ quark, and $\pi/K$ decay (see Fig.~\ref{fig:ptrel})
as determined from {\footnotesize ISAJET} MC.
The shape for $\pi/K$ decays was found to agree with the data 
distribution sample in the $p_T^{\mu}$ range 0.5--1.0 GeV/$c$
which is dominated by these decays.
As is shown in Fig.~\ref{fig:b_fracts}, the NLO QCD Monte Carlo agrees quite
well with the measured $f_b$ obtained in the $p_T^{\rm rel}$ fits of both the
entire data sample, and the subset of events from the cross section sample that
have a jet associated with a muon. Having shown that the MC is reliable for 
events with muons with jets, we assume it is also reliable for inclusive muons.

\begin{figure}[t]
\epsfxsize=7in\epsfbox{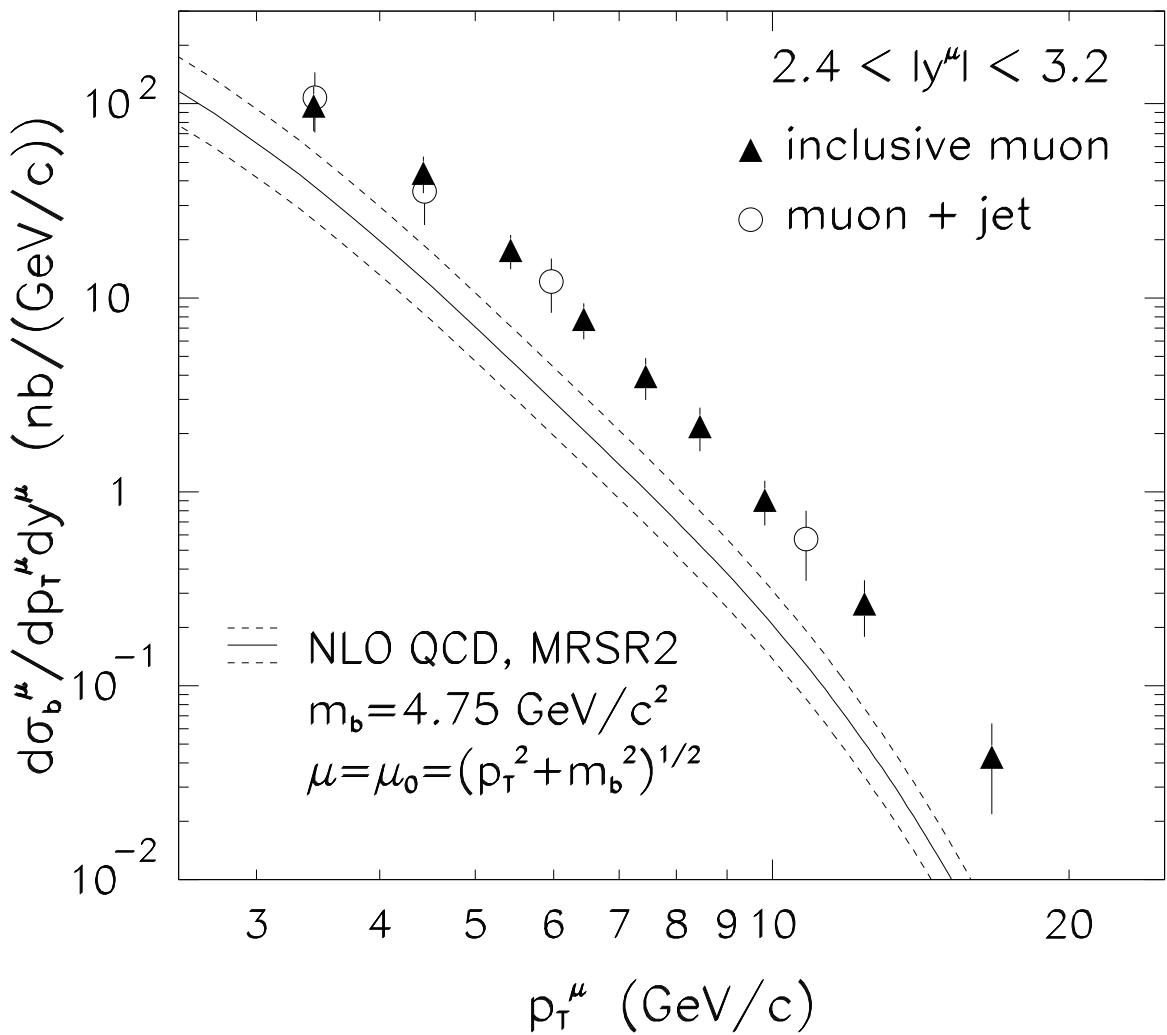}
\caption{The cross section for muons from $b$ quark decay
as a function of $p_T^{\mu}$ (per unit rapidity) as measured
with the inclusive muon sample (triangles) and its sub-sample of events 
that have a jet associated with the muon (circles).
The solid curve is the NLO QCD
prediction, with the dashed curves 
representing the theoretical uncertainties.}
\label{fig:bmu_xs}
\end{figure}

Subtracting the $\pi/K$ contribution from the inclusive
muon cross section and multiplying the result by the QCD MC predictions for $f_b$ 
gives the cross section for muons originating from $b$ quark decay. 
Our measurement, which includes
both muon charges, and sequential 
$b \rightarrow c \rightarrow \mu$ decays, is shown in Fig.~\ref{fig:bmu_xs} and
Table~\ref{tab:bmu_xs}.

The systematic uncertainties of this measurement 
include those of the inclusive muon cross section,
with additional uncertainties due to $f_b$ and the $\pi/K$ subtraction.
The contribution to the muon cross section from $\pi/K$ decay is predominantly
in the low $p_T^{\mu}$ bins. Conservatively assuming that the data in the 2 -- 3~GeV/$c$
bin (see Fig.~\ref{fig:mu_xs}) is entirely due to $\pi/K$ decay, we determine that the
{\footnotesize ISAJET} normalization is correct to within a factor of 1.35. This factor
is used to determine the uncertainty in the higher $p_T^{\mu}$ bins. 

Also shown in the figure is a cross check of our measurement. We determine the
cross section using the same events, but now require the muon to be associated 
with a jet, and use the values for $f_b$ that were determined in the $p_T^{\rm rel}$
fits to the entire data sample. We obtain the same cross section (within statistical
errors) as we do in the inclusive muon analysis. 

The NLO QCD predictions for the forward muon cross section from $b$ quark decay
are also shown in Fig.~\ref{fig:bmu_xs} as a function of $p_T^{\mu}$. They
match the shape of the measured cross section fairly well, but are
approximately a factor of four lower than the data.   

\begin{figure}[t]
\epsfxsize=7in\epsfbox{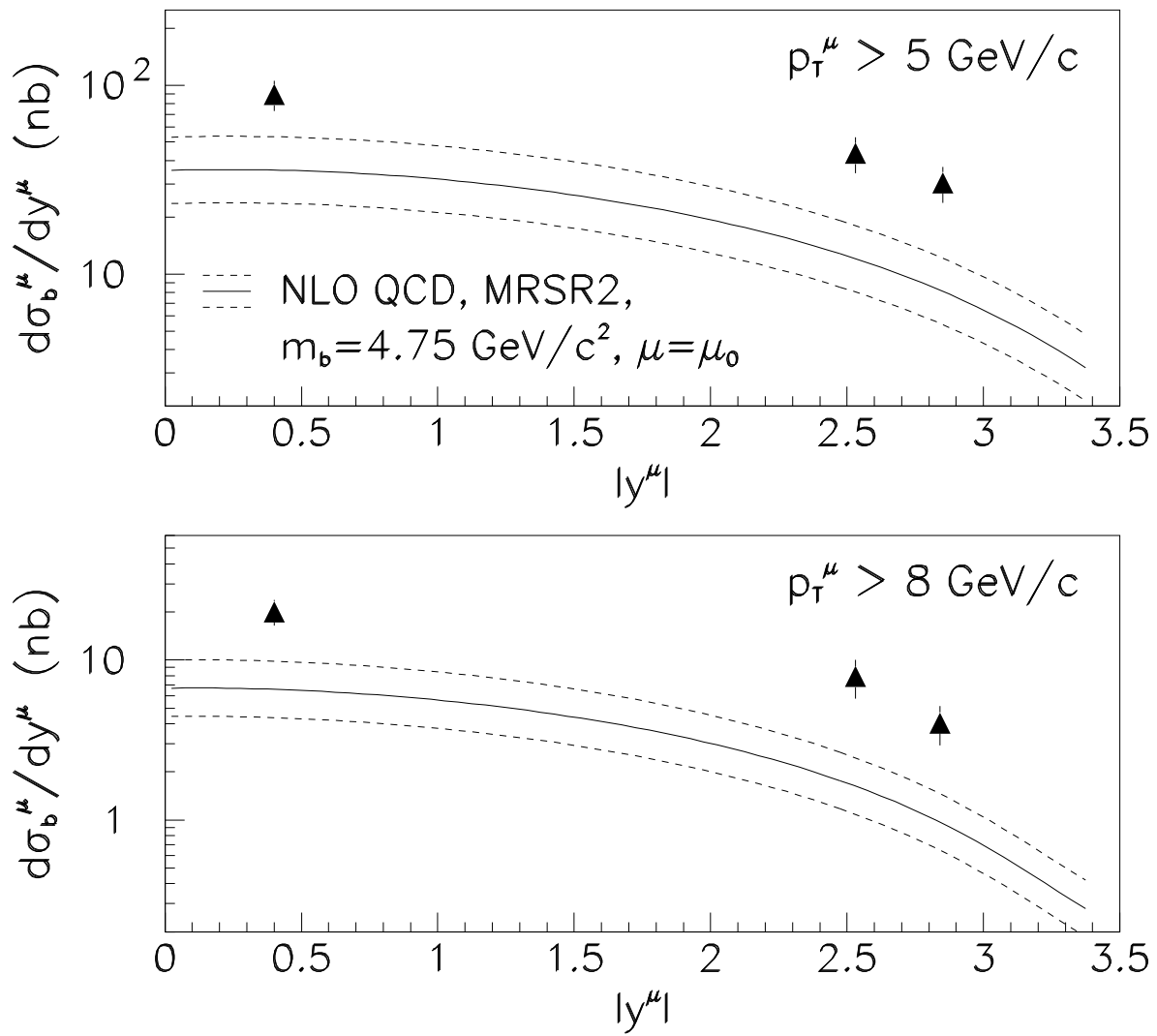}
\caption{
The cross section of muons from $b$ quark 
decay as a function of $|{\rm \sl y}^{\mu}|$ for $p_T^{\mu}> 5$ GeV/$c$, 
and $p_T^{\mu}> 8$ GeV/$c$. The solid curves are the NLO QCD predictions,
with uncertainty bands shown by the 
dashed lines.}
\label{fig:bmu_dxdy}
\end{figure}

By combining the forward cross section with that
of a previous D\O\ measurement in the central rapidity range ($|{\rm \sl y}^{\mu}|<0.8$)
\cite{ref:DO_bx}
we can study the rapidity dependence of $b$ quark production. 
Our measurement of the cross section for muons from $b$ quark 
decay as a function of rapidity ($d \sigma_{b}^{\mu}/d|{\rm \sl y}^{\mu}|$) is
shown in Fig.~\ref{fig:bmu_dxdy} for both $p_T^{\mu}> 5$ GeV/$c$
and $p_T^{\mu}>8$ GeV/$c$. The ratios between data and theory  are shown in 
Table~\ref{tab:xs_ratio}. We find that next-to-leading 
order QCD calculations do not reproduce the measurements.

\newcolumntype{d}{D{.}{.}{2.4}}
\newcolumntype{e}{D{.}{.}{3.1}}
\newcolumntype{f}{D{.}{.}{2.2}}
\newcolumntype{g}{D{.}{.}{3.3}}
\newcolumntype{h}{D{.}{.}{2.1}}
\newcolumntype{i}{D{.}{.}{1.1}}
\begin{table}[ti]

\centering
\caption{The cross section of muons from $b$ quark decay 
compared to NLO QCD. Errors are
statistical and systematic added in quadrature.}  
\vspace{0.5cm}
\label{tab:xs_ratio}
\begin{tabular}{cdecfghci}
\hline
\hline
\\
\multicolumn{2}{l}{$p_T^{\mu} > 5$ GeV/$c$} \\ 
\hline
         &       & \multicolumn{3}{c}{measured}  & \multicolumn{1}{c}{theory} & \\
rapidity & \multicolumn{1}{c}{$\langle y \rangle$} & \multicolumn{3}{c}{$\sigma^{\mu}_b$ (nb)} &
\multicolumn{1}{c}{$\sigma^{\mu}_b$ (nb)} & \multicolumn{3}{c}{ratio} \\ 
\hline
0.00 -- 0.80     & 0.40   &  89   & $\pm$ & 16   &  36     & 2.5 & $\pm$ & 0.4      \\
2.40 -- 2.65     & 2.53   &  43.5 & $\pm$ & 9.4  &  12     & 3.6 & $\pm$ & 0.8      \\
2.65 -- 3.20     & 2.85   &  30.5 & $\pm$ & 6.6  &  8.4    & 3.6 & $\pm$ & 0.8      \\
\hline
\\

\multicolumn{2}{l}{$p_T^{\mu} > 8$ GeV/$c$} \\ 
\hline
         &       & \multicolumn{3}{c}{measured}  & \multicolumn{1}{c}{theory} & \\
rapidity & \multicolumn{1}{c}{$ \langle y \rangle$} & \multicolumn{3}{c}{$\sigma^{\mu}_b$ (nb)} &
\multicolumn{1}{c}{$\sigma^{\mu}_b$ (nb)} & \multicolumn{3}{c}{ratio} \\ 
\hline
0.00 -- 0.80    & 0.40   &  20.1 & $\pm$ & 3.7  &  6.6     & 3.0 & $\pm$ & 0.6      \\
2.40 -- 2.65    & 2.53   &  7.9  & $\pm$ & 2.2  &  1.6     & 4.8 & $\pm$ & 1.3      \\
2.65 -- 3.20    & 2.84   &  4.1  & $\pm$ & 1.1  &  0.99    & 4.0 & $\pm$ & 1.1    \\
\hline
\hline
\end{tabular}
\end{table}

There have been some recent theoretical attempts to account for the discrepancy
between data and theory.
New calculations based on a variable flavor number
scheme~\cite{ref:Olness} predict an increase in the $b$ quark
cross section by a factor of 1.2 -- 1.5 with respect to the 
standard calculations which use a fixed flavor  number scheme. An increase in the
$B$-meson cross section of 50\% in the forward region and 30\% in
the central region can also be obtained by using a stiffer $b$ quark
fragmentation function than the standard Peterson form~\cite{ref:Mangano}.
Neither of these effects, however, can bring the predicted cross sections up
to the measured values.

In summary, we have measured the inclusive muon cross section, and the
cross section for
muons originating from $b$ quark decay, in the forward rapidity
region of $2.4 < | {\rm \sl y}^{\mu} | < 3.2$. We find that
next-to-leading order QCD calculations underestimate $b$ quark production
by a factor of four in this region.

%
We thank the Fermilab and collaborating institution staffs for 
contributions to this work, and acknowledge support from the 
Department of Energy and National Science Foundation (USA),  
Commissariat  \` a L'Energie Atomique (France), 
Ministry for Science and Technology and Ministry for Atomic 
   Energy (Russia),
CAPES and CNPq (Brazil),
Departments of Atomic Energy and Science and Education (India),
Colciencias (Colombia),
CONACyT (Mexico),
Ministry of Education and KOSEF (Korea),
and CONICET and UBACyT (Argentina).


\begin{references}

\bibitem{ref:DO_bx} S.~Abachi {\it et al.} (D\O\ Collaboration),
    Phys. Rev. Lett. {\bf 74}, 3548 (1995).

\bibitem{ref:DO_dim}  B.~Abbott {\it et al.} (D\O\ Collaboration),
       FERMILAB-PUB-99/144-E, to be submitted to Phys. Rev. D.    

\bibitem{ref:CDF_bx} F.~Abe {\it et al.} (CDF Collaboration),
     Phys. Rev. Lett. {\bf 71}, 500, 2396, 2537 (1993);
     Phys. Rev. Lett. {\bf 75}, 1451 (1995).

\bibitem{ref:MNR}
P.~Nason, S.~Dawson and R.K.~Ellis, {Nucl. Phys.}
{\bf B327}, 49 (1989);
M.~Beenakker {\it et al.} {Nucl. Phys.}
{\bf B351}, 507 (1991); 
M.~Mangano, P.~Nason and G. Ridolfi, {Nucl. Phys.}
{\bf B373}, 295 (1992).

\bibitem{ref:CDF_corr} F.~Abe {\it et al.} (CDF Collaboration),
     Phys. Rev. D {\bf 53}, 1051 (1996);
    Phys. Rev. D {\bf 55}, 2546 (1997).

\bibitem{ref:CDF_forw} F. Abe {\it et al.} (CDF Collaboration), 
       FERMILAB-PUB-98/392-E, submitted to Phys. Rev. D. 

\bibitem{ref:D0_det} S. Abachi {\it et al.} (D\O\ Collaboration),  
Nucl. Instrum. Methods Phys. Res., A {\bf 338}, 185 (1994).

\bibitem{ref:sam_det} C. Brown {\it et al.} (D\O\ Collaboration),
  Nucl. Instrum. Methods Phys. Res., A {\bf 279}, 331 (1989);
  Yu. Antipov {\it et al.}, Nucl. Instrum. Methods Phys. Res., 
  A {\bf 297}, 121 (1990).

\bibitem{ref:sampsi} B.~Abbott {\it et al.} (D\O\ Collaboration),
    Phys. Rev. Lett. {\bf 82}, 35 (1999).
 
\bibitem{ref:GEANT} R.~Brun and F.~Carminati, CERN Program
Library Long Writeup W5103, 1993 (unpublished).

\bibitem{ref:ISAJET} F. Paige and S.D. Protopopescu, BNL Report
                BNL-38034, 1986 (unpublished), release V7.22.    

\bibitem{ref:CDF_pik} F. Abe {\it et al.} (CDF Collaboration),
 Phys. Rev. Lett. \
                {\bf 61}, 1819 (1988).


\bibitem{ref:MRS} A.~Martin, R.~Roberts and  J.W.~Stirling,
    Phys. Rev. D {\bf 47}, 867 (1993); {\bf 51}, 4756 (1995).

\bibitem{ref:Peterson} C.~Peterson {\it et al.},
     Z. Phys. C {\bf 36}, 163 (1987).

\bibitem{ref:PDG} C.~Caso {\it et al.},
     Euro. Phys. J. {\bf C3}, 1 (1998). 

\bibitem{ref:Olness} F.~Olness, R.~Scalise and Wu-Ki Tung, 
    Phys. Rev. D {\bf 59}, 14506 (1999).

\bibitem{ref:Mangano} M.~Mangano, CERN preprint TH-97-328, 1997 (unpublished).

\end{references}
\end{document}